\def\cm{{\rm\thinspace cm}}

\def\erg{{\rm\thinspace erg}}

\def\K{{\rm\thinspace K}}
\def\keV{{\rm\thinspace keV}}
\def\km{{\rm\thinspace km}}

\def\Lsun{\hbox{$\rm\thinspace L_{\odot}$}}

\def\Mpc{{\rm\thinspace Mpc}}
\def\Msun{\hbox{$\rm\thinspace M_{\odot}$}}
\def\pc{{\rm\thinspace pc}}

\def\s{{\rm\thinspace s}}
\def\yr{{\rm\thinspace yr}}
\def\sr{{\rm\thinspace sr}}

\def\ergpcmsqps{\hbox{$\erg\cm^{-2}\s^{-1}\,$}}

\def\ergps{\hbox{$\erg\s^{-1}\,$}}

\def\kmps{\hbox{$\km\s^{-1}\,$}}

\def\pcmsq{\hbox{$\cm^{-2}\,$}}

\def\psqcm{\hbox{$\cm^{-2}\,$}}

\def\kmpspMpc{\hbox{$\kmps\Mpc^{-1}$}}
\documentstyle[psfig]{mn}
\begin{document}
\title{The obscured growth of massive black holes}
\author[]
{\parbox[]{6.in} {A.C.~Fabian\\
\footnotesize
Institute of Astronomy, Madingley Road, Cambridge CB3 0HA \\}}

\maketitle
\begin{abstract}
The mass density of massive black holes observed locally is consistent
with the hard X-ray Background provided that most of the radiation
produced during their growth was absorbed by surrounding gas. A simple
model is proposed here for the formation of galaxy bulges and central
black holes in which young spheroidal galaxies have a significant
distributed component of cold dusty clouds which accounts for the
absorption. The central accreting black hole is assumed to emit both a
quasar-like spectrum, which is absorbed by the surrounding gas, and a
slow wind. The power in both is less than the Eddington limit for the
black hole. The wind however exerts the most force on the gas and, as
earlier suggested by Silk \& Rees, when the black hole reaches a
critical mass, it is powerful enough to eject the cold gas from the
galaxy, so terminating the growth of both black hole and galaxy. In
the present model this point occurs when the Thomson depth in the
surrounding gas has dropped to about unity and results in the mass of
the black hole being proportional to the mass of the spheroid, with
the normalization agreeing with that found for local galaxies by
Magorrian et al. for reasonable wind parameters. The model predicts a
new population of hard X-ray and sub-mm sources at redshifts above one
which are powered by black holes in their main growth phase.

\end{abstract}


\section{Introduction}

The local mass density of massive black holes residing in the nuclei
of galaxies (Magorrian et al 1998; Richstone et al 1998) is in good
agreement with that expected from the intensity of the X-ray
background at 30~keV (Fabian \& Iwasawa 1999; Salucci et al 1999). The
X-ray background is assumed to be due to radiatively-efficient
accretion onto black holes at redshifts $z\sim 1-2$, with accretion
accounting for the bulk of their mass. An important requirement is the
presence of large column densities of absorbing matter, $N_{\rm H}\sim
10^{24}\pcmsq$, around many of the X-ray sources. Intrinsically they
are assumed to radiate the broad-band spectrum of a typical quasar
(Elvis et al 1994), with a large UV bump, and a power-law X-ray
continuum of photon index, $\Gamma\approx2$, up to a few $100\keV$.
Photoelectric absorption in this matter then hardens the observed
spectrum considerably below 30~keV so that the cumulative spectrum
from a population of sources with a range of column densities and
redshifts resembles that of the observed X-ray Background (Setti \&
Woltjer 1989; Madau et al 1994; Celotti et al 1995; Matt \& Fabian 1994;
Comastri et al 1995; Wilman \& Fabian 1999). Provided that the sources
are not too Thomson-thick, it is only above about 30~keV that the
unabsorbed radiation from growing black holes is observed.

Two major issues for the model are discussed here; a) the typical mass
and luminosity of the obscured objects and b) the source of the
obscuring matter. The first a) is an issue because few luminous
obscured quasars $L(2-10\keV)>10^{44}\ergps$ have yet been found by
X-ray (or other waveband) surveys or studies (Boyle et al 1998;
Brandt et al 1997; Halpern et al 1999), so it might seem that highly
obscured objects are only of low luminosity. This could be a serious
problem for the growth of black holes of mass $10^9\Msun$ or more,
which might have luminosities considerably more than that. The second
b) requires large column densities to obscure most of the Sky as seen
from the sources, in order that most accretion power in the Universe
is absorbed (Fabian \& Iwasawa 1999 estimate that about 85 per cent is
absorbed). We have previously argued that a circumnuclear starburst is
responsible for the obscuration in nearby objects (Fabian et al 1998),
but in its simple form could be difficult to sustain if the central
black hole takes a Gyr or more to grow to its present mass.

A model is presented here in which a black hole grows with its
surrounding stellar spheroid by the accretion of hot gas. The gas is
also cooling and forming a distributed gaseous cold component within
which stars slowly form. In the central regions the cold component
provides the required column density for absorption of the quasar
radiation from the accreting black hole. The quasar is assumed to make
a slow wind, which eventually becomes powerful enough to blow away the
absorbing gas, probably due to a wind, and an unobscured quasar is
seen (see Silk \& Rees 1998 and Blandford 1999 for a similar wind
model). Because the fuel supply for the black hole has also been
ejected however, the quasar dies when its accretion disc is exhausted.
The high obscuration phase occurs only during the growing phase and
thus at redshifts greater than unity which have not been explored yet
in the hard X-ray band. Reprocessing of the radiation by the absorbing
gas into far infrared emission may make these objects detectable in
the sub-millimetre band.

\section{Growth of the black hole}

Magorrian et al (1998) find from the demographics of nearby massive
black holes that the mass of the hole $M_{\rm BH}$ is proportional to
the mass of the surrounding spheroid $M_{\rm sph}$ or bulge; $M_{\rm
BH}\approx 0.005M_{\rm sph}$. There is considerable scatter about this
relation of order $\pm1$~dex. This can be combined with the mass
function of galactic bulges, where most of the mass resides. Salucci et
al (1999) use the Schechter function luminosity function of bulges to
obtain the local mass function of black holes. Most of the mass lies
in black holes of individual mass around the break in the function,
i.e. $3\times 10^8\Msun$.

Taking this as a typical mass, and assuming growth by accretion with a
radiative efficiency of $\eta=0.1 \eta_{-1}$, the bolometric
luminosity of the final object $L_{\rm B}=3\times 10^{46}-3\times
10^{44}\ergps$ if it is at one and 0.01 times the Eddington limit
$L_{\rm Edd}$, respectively. The Salpeter time for the growth of the
black hole, i.e. the mass doubling time, $t_{\rm s}=3\times 10^7
(L_{\rm Edd}/L_{\rm B})\eta_{0.1}\yr.$ If we now assume that the
initial mass of all massive black holes is less than that of the
central black hole in our Galaxy, say $\sim 10^6\Msun$, and has grown
to $3\times 10^6\Msun$ within 3~Gyr, which corresponds to $z\sim2$,
then $8t_{\rm s}<3\times 10^9\yr$ and the black hole must have grown
at a rate of at about 10 per cent of the Eddington rate or more. Thus
over the final $3\times 10^8\yr,$ $L_{\rm B}>3\times 10^{45}\ergps$.
From the work of Elvis et al (1994) the observed 2--10~keV luminosity
is about 3 per cent of the bolometric one for a quasar so we can
conclude that $L(2-10\keV)>10^{44}\ergps$ during the growth of a
typical black hole and we are dealing with powerful, obscured quasars.

It is clear that the objects making up the X-ray background, and the
radiative growth phase of most of the mass in nearby black holes are
not represented by any object observed so far. 

\section{The obscuration}

The spectrum of the X-ray Background requires that most accretion
power is absorbed. This means that some absorption (say $N_{\rm
H}>10^{22}\psqcm$) occurs in 90 per cent of objects and heavy
absorption ($N_{\rm H}>10^{24}\psqcm$) in 30--50 per cent of them. 10
per cent are unabsorbed and are the quasars identified in blue optical
surveys or are bright in the soft X-ray band of ROSAT. So far these
could be distinct different populations of quasars or different phases
in the growth of all quasars.

Such high absorbed fractions mean that the covering fraction of the
sky by high column density material seen from a growing black hole
must approach $4\pi \sr$. This cannot be provided by any thin disc or
by the standard torus of unified models. The absorbing matter is
probably concentrated within the innermost few 100~kpc, or its total
mass becomes high. It must presumably be cold and fairly neutral, or
it will not absorb the X-rays. At first sight this is at variance with
it being space covering, especially if it must be so for a Gyr or
more. Such matter should collide and dissipate into a disclike
structure.

Following our earlier work on a circumnuclear starburst (Fabian et al
1998), it is plausible that collisions do take place leading to
dissipation but that some massive star formation occurs in the shocked
and cooled dense gas. The winds and supernovae from those stars then
supply the energy to keep the rest of the cold matter in a chaotic and
space covering state.

A more detailed model can be developed by assuming that the black hole
is growing at the same time as the galaxy does. The stellar spheroid
continues to grow by cooling of the gas heated by gravitational
collapse of the protogalaxy. It is likely that the hot phase density
while the galaxy continues to grow is the maximum possible, which
means that the radiative cooling time of the gas equals the
gravitational infall time. This condition has been studied in the
context of quasar fuelling and growth by Nulsen \& Fabian (1999). The
gas accretes in a Bondi flow, probably forming a disc well within the
Bondi radius. The accretion rate is such that it can typically be 10
per cent of the Eddington value.

The situation as envisaged here is essentially a maximal cooling flow
and we can use the properties of observed cooling flows (Fabian 1994
and references therein) to indicate how the cooled gas is distributed
and how it may lead to absorption. X-ray observations of cluster
cooling flows show that the mass deposited by cooling is distributed
with $\dot M(<r)\propto r$, where $r$ is the radius. The density
distribution of the cooled gas is therefore $\rho\propto r^{-2}$. If
the gravitational potential of the galaxy is isothermal it remains so.
X-ray absorption with column densities of the order of $10^{21}\psqcm$
is also observed in many cluster cooling flows. Although not observed
in other wavebands, it could represent very cold, dusty gas (Fabian et
al 1994). 

It is therefore proposed that the cooled gas in protogalaxies does not
all rapidly form stars but that much of it forms long-lived cold dusty
clouds. As discussed above, energy from the massive stars helps to
prevent the clouds rapidly dissipating into a large disc. The inner
regions of protogalaxies are therefore highly obscured. 

The column density through the cold clouds is obtained from
integration of the cold gas density distribution, assumed to be
$$n=n_0r_0^2r^{-2}.\eqno(1)$$ If the cold clouds are a fraction $f$ of the
total mass within radius $r, M(<r)=2v^2r/G,$ then
$$n_0r_0^2=f{v^2\over{2\pi G m_{\rm p}}}\eqno(2)$$ and the column density in
to radius $r_{\rm in}$ $$N_{\rm H}={v^2\over{2\pi G m_{\rm p}}}{f\over
r_{\rm in}} =10^{24}f{v_{2.5}^2\over r_2}
\psqcm.\eqno(3)$$ Here the (line of sight) velocity dispersion of the (isothermal) 
spheroid is
$300v_{2.5}\kmps$ and $r_{\rm in}=100r_2\pc.$ It is therefore
reasonable to assume that column densities at the level required by
X-ray Background models can be produced in this manner, provided that
$f$ is fairly high.

It is interesting that the mass within the radius where the column
density becomes Thomson thick ($N_{\rm T}=\sigma_T^{-1}\sim 2\times
10^{24}\psqcm,$ where $\sigma_{\rm T}$ is the Thomson electron
scattering cross section, is given by $$r_{\rm T}={v^2f\over{2 \pi G
m_{\rm p} N_{\rm T}}},\eqno(4)$$ for which the enclosed mass $$M_{\rm
T}={v^4f\over{2 \pi G^2 m_{\rm p} N_{\rm T}}}.\eqno(5)$$ $M_{\rm T}$
is thus approximately proportional to the mass of the whole spheroid,
$M_{\rm sph}$, since approximately $M_{\rm sph}\propto v^4$, as
suggested by the Faber-Jackson relation amd by galaxy modelling
(Salucci \& Persic 1997).  The last authors note that $r_{\rm
e}\propto L_B^{0.7}$ and $M_{\rm sph}\propto L_B^{1.35},$ which with
$v\propto (M/r)^{0.5}$ yields a result very close to $M_{\rm
sph}\propto v^4$.

As the stellar content of the galaxy and the central black hole grow,
the accretion radius of the hole also grows. When $M_{\rm BH}\sim
M_{\rm T}$ the accretion radius is at $r_{\rm T}$ and the Thomson
depth of the obscuring matter around the quasar has dropped to unity. 

The growth of the quasar continues until it runs out of gas. This
could be because all the gas has cooled and formed stars. More likely
it is connected with the central quasar (Silk \& Rees 1998; Blandford
1999). Consider the possibility that as well as radiation the quasar
produces a wind at velocity $v_{\rm W}\ll c$ and with a kinetic power
$L_{\rm w}$ which scales with the radiated power $L_{\rm rad}$ as
$L_{\rm w}/L_{\rm rad}=L_{\rm Bol}/L_{\rm Edd}$. Thus when the
bolometric power is 10 per cent of the Eddington value the power of
the wind is 10 per cent of the radiated power. Note that winds and
outflows are observed from many classes of accreting objects (see e.g.
Livio 1997).

The ratio of wind to (optically-thin) radiation pressure is then
$${P_{\rm w}\over P_{\rm rad}}={1\over 2}{L_{\rm w}\over L_{\rm
rad}}{c\over v_{\rm w}}={1\over 2}{L_{\rm Bol}\over L_{\rm
Edd}}{c\over v_{\rm w}}.\eqno(6)$$ For an optically-thin medium of Thomson
depth $\tau_{\rm T}$, the effective Eddington limit $$L^{\rm w}_{\rm
Edd}=L_{\rm Edd}{\tau\over 2}{v_{\rm w}\over c}.\eqno(7)$$ This means that if
$v_{\rm w}\sim {\rm 15,000}\kmps$ and $L_{\rm Bol}/L_{\rm Edd}\sim 0.1$ then
the wind pressure balances gravity. A wind force slightly in excess of
this can therefore eject the gas. Since $\tau \propto r^{-1}$ and
$M\propto r$, this condition applies throughout the galaxy and {\it
all} the gas is ejected. (The assumption here that the mass is in a
thin shell is applicable to the gas as it is swept up.)

This result can be placed on a surer footing by noting that the force
balance between gravity acting on a column of matter at radius $r$ of
total mass $N_{\rm H} 4\pi r^2 m_{\rm p}$ and the outward force due to
a wind $2 L_{\rm w}/v$ yields the limiting luminosity $$L^{\rm w}_{\rm
Edd}=2\pi GMm_{\rm p}N_{\rm H} v_{\rm w}.\eqno(8)$$ Substituting for $M$
and $N_{\rm H}$ in the model galaxy we obtain $$L^{\rm w}_{\rm
Edd}={{{2v^4 f v_{\rm w}}\over{G}}}.\eqno(9)$$ The gas is ejected by the wind
when the wind power exceeds this limit, i.e. when $$L_{\rm w}>L^{\rm
w}_{\rm Edd}.\eqno(10)$$ If $L_{\rm w}=a L_{\rm Edd}$ then the limit occurs
when $$a {{4 \pi G M m_{\rm p} c}\over \sigma_{\rm T}}= {{v^4 f v_{\rm
w}}\over{2G}},\eqno(11)$$ or at the critical mass $$M_{\rm c}={v^4\over{2\pi
G^2 m_{\rm p}N_{\rm T}}}{v_{\rm w}\over c}{f\over a}={M_{\rm T}\over
2a}{v_{\rm w}\over c}.\eqno(12)$$ Thus if $v_{\rm w}\sim 0.1 c$, and $a\sim
0.1$ then $M_{\rm c}\sim M_{\rm T}/2.$ Assuming that most of the mass
within $r_{\rm T}$ lies in the central black hole again means that
$M_{\rm BH}\sim M_{\rm T}\propto M_{\rm sph}.$ Most of the power
during the main growth phase of a massive black hole is then radiated
into gas with a Thomson depth of about unity, as required for
modelling the X-ray Background spectrum.

Clearly $f$ cannot equal unity if a galaxy is to be formed. In
practice it will be a function of time. The important issue here is
that it cannot be small, i.e. it probably lies in the range of 0.1 --
0.5.

The black hole mass is therefore proportional to the spheroid mass, as
observed (Magorrian et al 1998). The quasar is now unobscured and is
observable as an ordinary blue excess object for as long as it has
fuel. A reasonable estimate would be about a million yr for the disc to
empty. The ejection phase is tentatively identified with broad
absorption line quasars (BAL quasars; see e.g. Weymann 1997).

The normalization for the relation between $M_{\rm BH}$  and $M_{\rm
sph}$ is obtained by using the Faber-Jackson relation given by Binney
\& Tremaine (1992), where $v=220(L/L_\star)^{0.25}\kmps$ and
$L_\star=4\times 10^{10}\Lsun$ in the V-band, and a mass-to-light
ratio in that band of 6 (both for a Hubble constant of $50\kmpspMpc$).
The result is $${M_{\rm T}\over M_\star} =0.01\eqno(13) $$ so
$${M_{\rm BH}\over M_{\rm sph}}\approx 0.005,\eqno(14)$$ for the
values of $a$ and $v_{\rm w}$ above. This is in good agreement with
the relation found by Magorrian et al (1998). In detail, the weak
variations in $M/L$ such as summarized by Salucci \& Persic (1997)
could shift the observed relation from being strictly linear. 

Note that the ordinary quasar phase markes the end of the main growth
phase of both the black hole and its host spheroid. The central black
hole in galaxies therefore has a profound effect on the whole galaxy,
in providing a limit to the stellar component. In this context, Silk
\& Rees (1998) have shown that a powerful quasar wind could end the
formation of a galaxy and Blandford (1999) has noted that it could
prevent the formation of a galactic disc.

\section{Discussion}

It has been shown here that a black hole, growing by accretion in the
centre of a young, isothermal, spheroidal galaxy which has a large
mass component of cold gas, will appear highly obscured ($\tau_{\rm
T}>1$). Provided that a significant fraction of the accretion power,
assumed to be close to the Eddington limit for the black hole, is
released as a sub-relativistic wind, then, as suggested by Silk \&
Rees (1998) the cold, and accompanying hot, gas components are ejected
from the galaxy when the Thomson depth to the outside $\tau_{\rm T}$
drops to about unity. The central accretion source then appears as an
unobscured quasar which lasts until its disc empties. The growth of
both the central black hole and the stellar body of the galaxy then
terminate, unless fresh gas and stars are brought in from outside, for
example by a merger. Assuming that most of the mass within the region
where $\tau_{\rm T}\sim 1$ has been accreted into the black hole, it
is found that its mass $M_{\rm BH}\propto M_{\rm sph}$, the mass of
the spheroid of the galaxy.

The rough explanation for the proportionality is that a more massive
galaxy has more cold gas and so requires a more powerful wind to eject
the gas. A stronger wind requires a more massive black hole. The
normalization results by relating the wind power to the Eddington
limit.

The model accounts for the bulk of the X-ray Background which requires
that most accretion power, resulting from the growth of massive black
holes, is obscured. Also, such obscured accretion leads to agreement
with the local mass density of black holes (Fabian \& Iwasawa 1999).
Most of the absorbed radiation will be reradiated in the far
infrared/sub-mm bands and contribute to the source counts and
backgrounds there. It can plausibly account for some of the sub-mm
sources recently discovered by SCUBA (Barger et al 1998; Hughes et al
1998, Blain et al 1999). Indeed it predicts a population of distant
Ultra-Luminous Infrared Galaxies (ULIRGs; Sanders \& Mirabel 1996)
associated with the main growth phase of massive black holes and
distinct from the nearby population, which is probably due to mergers
briefly fuelling central starbursts and fully grown black holes.
(Whether radiation from the black hole can then dominate the ULIRG
emission depends on the Eddington limit, i.e. the mass of the black
hole, relative to the strength of the starburst; see e.g. Wilman et al
1999.) The bolometric power of a typical growing black hole must be
high, $L_{\rm Bol}>10^{46}\ergps$; which means an observed 2--10~keV
luminosity when $N_{\rm H}\sim 2\times 10^{24}\psqcm$ of about
$3\times 10^{-15}\ergpcmsqps$. Current ASCA (e.g. Ueda et al 1998) and
BeppoSAX (Fiore et al 1999) hard X-ray surveys have not probed deep
enough to reveal these objects, although they should be obvious in the
deeper surveys planned for Chandra and XMM. These surveys will enable
the major growth phase of black holes and, with optical/infrared
identifications, their redshift distribution to be studied.

Note that the obscured growth phase of a massive black hole represents
a distinctly different phase from the briefer unobscured phase
predicted after the gas is ejected and the quasar dies or any later
phase, obscured or unobscured, when the quasar is revived by a merger
or other transient fuelling event. These last phases are the ones
which have been observed so far; the major growth phase has not. It is
unlikely therefore that the properties of the major growth phase are
obtainable by any simple extrapolation from observations of any
transient recent phases, i.e. from studies of the properties of active
galaxies at low redshift ($z<0.5$).

It is important that the cold gas in the young galaxy, which provides
the X-ray (and other waveband) obscuration, be metal enriched, and
probably dusty. This is likely to be a consequence of continued star
formation, particularly of massive stars, throughout the galaxy. The
energy of stellar winds and supernovae help to keep the cold gas space
covering. Note that the injected metals and stellar mass loss will be
distributed both as assumed for the stars and the cold gas. Indeed the
metallicity of the obscuring gas closest to the black hole may be
higher than the solar values, which leads to better fits to the X-ray
Background spectrum (Wilman \& Fabian 1999). This also accounts for
the high metallicity inferred in the broad-line region gas for many
quasars (Hamann \& Ferland 1999).

The column density distribution of the gas obscuring the radiation
from the growing black hole will be such that most power is emitted
just before the gas is ejected, which happens around $\tau_{\rm T}\sim
1$. The fraction of the power radiated at optical depths greater than
$\tau_{\rm T}$ scales roughly as $\tau_{\rm T}^{-1}$. Angular momentum
and other factors may cause the gas in the young spheroidal galaxy,
and the wind, to not be completely spherical. Thus when the wind
ejects the gas it may do so most along one axis and only later eject
gas in other directions (if at all). This means that the column
density distribution for $\tau_{\rm T}<1$ is complicated to predict
and is best found from the shape of the X-ray Background spectrum.

An important consequence for the stellar radiation in young growing
galaxies is that most of it too is obscured. This agrees with recent
evidence for the star formation history of the Universe from sub-mm and
other observations. 

The ejection of the metal-rich cold gas in the young galaxy should
terminate its stellar growth, as well as growth of the black hole. The
final appearance of a galaxy is thus significantly affected by its
central black hole. How far the gas is ejected depends on how long the
unobscured quasar phase lasts and what the surrounding gas mass and
density is; whether for example the galaxy is in a group or cluster.
The most massive black holes will be in the most massive galaxies and
may last longest in the unobscured quasar phase. They might also be
surrounded by a hot intragroup medium which could prevent much of the
hotter space-filling phase from being ejected. If a surrounding hot
phase is a necessary ingredient for a radio source then such objects
might be more likely to be radio galaxies.

During the ejection phase the quasar might be classed as a BAL and
later it might be seen to be surrounded by extended metal-rich
filaments, depending on the velocity of ejection of the cold gas. The
metal-rich gas, if mixed with surrounding hot intracluster gas, will
enhance the local metallicity, providing one source for the extensive
metallicity gradients found by X-ray spectroscopy around many cD
galaxies in clusters (Fukazawa et al 1994). 

Finally, it is noted that the model requires a significant power
output in the form of a wind associated with the growth of black
holes. This wind power is dissipated as heat in the surrounding
medium. It may have a marked effect on surrounding intracluster gas
(Ensslin et al 1998; Wu, Fabian \& Nulsen 1999), possibly contributing
to the heating required to change the X-ray luminosity--temperature
relation, $L_{\rm x}\propto T_{\rm x}^\alpha$, from the predicted one
with $\alpha\sim 2$ to the observed one with $\alpha\sim 3.$ The
estimates of Wu et al (1999) indicate that it will also heat the
general intergalactic medium to a temperature of $\sim 10^7\K$ at
$z\sim 1-2$.

In summary, the growth of both massive black holes and galactic bulges
is a highly obscured, and related, process, best observed directly in
the hard X-ray band and indirectly, through radiation of the absorbed
energy, in the sub-mm band.

\section{Acknowledgements}

I thank the referee for comments and The Royal Society for support.

\end{document}